\documentstyle[12pt,epsfig,glas]{article}
\newlength{\dinwidth}                                                          
\newlength{\dinmargin}                                                         
\input epsf
\epsfverbosetrue
\begin{document}
\newcommand {\pom}  {I\hspace{-0.2em}P}
\newcommand {\xpom} {\mbox{$x_{_{\pom}}$}}
\newcommand {\Spom} {\mbox{$\Sigma_{_{\pom}}$}}
\newcommand{\sleq} {\raisebox{-.6ex}{${\textstyle\stackrel{<}{\sim}}$}}
\newcommand{\sgeq} {\raisebox{-.6ex}{${\textstyle\stackrel{>}{\sim}}$}}
\begin{titlepage}{GLAS-PPE/96--01}{\today}
\title{Diffraction at HERA: experimental perspective}
\author{Anthony T Doyle}
\begin{abstract}
Measurements of diffractive phenomena observed at HERA are reviewed. 
A short introduction to the theoretical background is presented.
The review focuses on the current experimental directions and discusses the 
exclusive production of vector mesons, the deep inelastic structure of
diffraction and complementary information from jet structures.
Emphasis is placed on the current sources of background and 
the experimental uncertainties.
\end{abstract}
\vspace{2cm}
\centerline{\em Talk presented at the Workshop on HERA Physics,}
\centerline{\em ``Proton, Photon and Pomeron Structure",}
\centerline{\em Durham, September 1995.}
\newpage
\end{titlepage}
\section{Introduction: maps of the pomeron}
At the last Durham workshop on HERA physics, HERA was heralded as
the new frontier for QCD. In the proceedings from that workshop, 
there was one 
theoretical contribution on ``Partons and QCD effects in
the pomeron"~\cite{ingelman}. 
Experimentally, the large rapidity gap events in deep inelastic scattering
were yet to be discovered and the first preliminary results on photoproduced 
vector mesons were just starting to appear.
Two years later, this workshop focuses on ``proton,
photon and pomeron structure": the inclusion of the word ``pomeron"
in the title of the workshop
reflects a series of diffractive measurements which have been made in the 
intervening period at HERA. This talk is therefore an opportunity 
to discuss ``{\it Measurements} of partons and QCD effects in the pomeron", 
based on results from the H1 and ZEUS collaborations.

The diffractive processes studied are of the form:
$$ e~(k)~+~ p~(P) \rightarrow e'~(k')~+~p'(P')~+~X,$$
where the the photon dissociates into the system $X$ and the outgoing 
proton, $p'$, remains intact, corresponding to single dissociation.
The measurements are made as a function of the photon virtuality,
$ Q^2 \equiv -q^2=-(k~-~k')^2, $
the centre-of-mass energy of the virtual-photon proton system,
$W^2=(q~+~P)^2$,
the mass of the dissociated system, $X$, denoted by $M^2$
and the four-momentum transfer at the proton vertex, given by
$t = (P~-~P')^2$.

The subject of diffraction is far from new: diffractive processes
have been measured and studied for more than thirty years~\cite{goul}. 
Their relation to the corresponding total cross sections at high energies
has been successfully interpreted via the introduction of a single pomeron 
trajectory with a characteristic $W^2$ and $t$ dependence~\cite{dl}.
The high-energy behaviour of the total cross sections is described by a
power-law dependence on $W^2$:
\begin{eqnarray}
\sigma \sim (W^2)^\epsilon
\end{eqnarray}
where $W$ is measured in GeV, 
$\epsilon = \alpha(0) - 1$ and $\alpha(0)$ is the pomeron intercept.
The slow rise of hadron-hadron total cross sections with increasing 
energy indicates that the value of $\epsilon \simeq 0.08$ i.e. the total
cross sections increase as $W^{0.16}$, although the latest $p\bar{p}$ 
data from CDF at $\sqrt{s}=1800$~GeV indicate 
$\epsilon \simeq 0.11$~\cite{CDF1}. 
The optical theorem relates
the total cross sections to the 
elastic, and hence diffractive, scattering amplitude at the same $W^2$:
\begin{eqnarray}
\frac{d\sigma}{dt} \sim (W^2)^{2(\epsilon-\alpha'\cdot |t|)}
\end{eqnarray}
where 
$\epsilon - \alpha' \cdot |t| = \alpha(t) - 1 $
and 
$\alpha'= 0.25~$GeV$^{-2}$ reflects the shrinkage of the
diffractive peak with increasing $W^2$.

Whilst these Regge-based models give a unified description of all
pre-HERA diffractive data, this approach is not fundamentally linked
to the underlying theory of QCD. 
It has been anticipated that at HERA energies
if either of the scales $Q^2$, $M^2$ or $t$ become larger than the 
QCD scale $\Lambda^2$, then 
it may be possible to apply perturbative QCD (pQCD) techniques, which 
predict changes to this power law behaviour, 
corresponding to an increase in the effective value of $\epsilon$
and a decrease of $\alpha'$.
This brings us from the regime of dominance of the slowly-rising
``soft" pomeron to the newly emergent ``hard" behaviour and the question
of how a transition may occur between the two.
Precisely where the Regge-based approach breaks down or 
where pQCD may be applicable is open to experimental question.
The emphasis is therefore on the internal (in)consistency of 
a wide range of measurements of diffractive and total cross sections.
As an experimentalist navigating around the various theoretical concepts of 
the pomeron, it is sometimes difficult to see which direction to take
and what transitions occur where~(Figure~1(a)).
However, from an experimental perspective, the directions are clear,
even if the map is far from complete~(Figure~1(b)).

\begin{figure}
\epsfxsize=10cm
\centering
\leavevmode
\epsfbox[45 497 507 732]{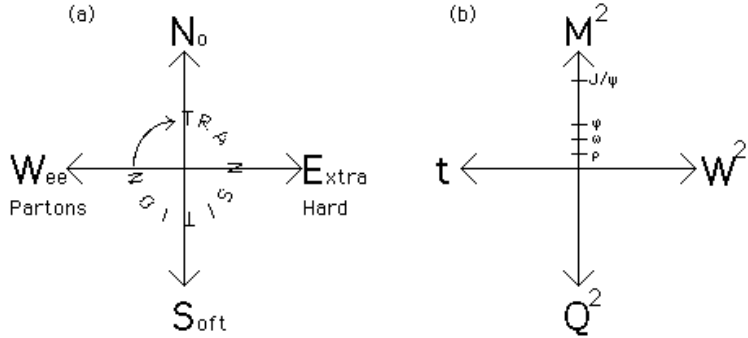}
\caption{Maps of the pomeron: (a) theoretical and (b) experimental directions.}
\end{figure}

The HERA collider allows us to observe a broad range of diffractive 
phenomena at the highest values of $W^2$.
What is new is that we have the ability to observe the variation 
of these cross sections at specific points on the $M^2$
scale, from the $\rho^0$ up to the $\Upsilon$ system as discussed in 
section~2.1. 
Similarly, the production cross section can be explored 
as a function of $Q^2$, using a virtual photon probe.
The high energy available provides a large rapidity span
of $\simeq$ 10 units ($\Delta(\eta) \sim \ln(W^2)$).
The observation of a significant fraction of events ($\simeq 10\%$)
with a large rapidity gap
between the outgoing proton, $p'$, and the rest of the final state, $X$, 
in deep inelastic scattering (DIS)
has led to measurements of the internal structure of the pomeron.
These results
are discussed in section 2.2.
Similar studies of events with high-$p_T$ jets and a large rapidity gap
have also been used to provide complementary information on this structure.
Also, the observation of rapidity gaps between jets, corresponding to 
large $t$ diffraction, are presented in section~2.3.
Finally, a first analysis of the leading proton spectrometer
data where the diffracted proton is directly measured is presented in
section~2.4.

\section{Signals and Backgrounds}
\subsection{Exclusive Production of Vector Mesons}
The experimental signals are the exclusive production of the vector mesons
in the following decay modes:
$$\rho^0\rightarrow \pi^+\pi^-~\cite{Hrho,Zrho,Hrho*,Zrho*}~~~~~~
\phi\rightarrow K^+K^-~\cite{Zphi,Zphi*}~~~~~~
J/\psi\rightarrow\mu^+\mu^-,e^+e^-~\cite{Hpsi,Zpsi,Hpsi*}.$$
First results on $\omega\rightarrow \pi^+\pi^-\pi^o$ and 
higher vector mesons ($\rho'\rightarrow \pi^+\pi^-\pi^o\pi^o$ and 
$\psi'\rightarrow\mu^+\mu^-,e^+e^-$) are in 
the early analysis stages and first candidates for 
$\Upsilon\rightarrow\mu^+\mu^-,e^+e^-$ are also appearing in the data.

The clean topology of these events results in typical errors on the measured 
quantities ($t$, $M^2$, $W^2$ and $Q^2$), reconstructed in the tracking 
chambers, of order 5\%.
Containment within the tracking chambers corresponds to a $W$ interval
in the range $40 \sleq W \sleq 140$~GeV. However, some analyses are restricted
to a reduced range of $W$ where the tracking and trigger systematics 
are well understood. Conversely, H1 have also used the 
shifted vertex data to
extend the analysis of the $\rho^0$ cross section to higher 
$<\!W\!> = 187$~GeV.
At small $t$ there are problems triggering and, to a lesser extent,
reconstructing the decay products of the vector meson. In particular,
the photoproduction of $\phi$ mesons is limited 
to $t \sgeq 0.1$~GeV$^2$, since the produced kaons are just above threshold and 
the available energy in the decay is limited.
In order to characterise the $t$-dependence, a fit to
the diffractive peak is performed. 
In the most straightforward approach, 
a single exponential fit to the $t$ distribution, 
$dN/d|t| \propto e^{-b|t|}$ for $|t|~\sleq~0.5~$GeV$^2$ is adopted. 

The contributions to the systematic uncertainties are similar in each 
of the measurements.
For example, the uncertainties on acceptance of photoproduced 
$\rho^0$'s are due to
uncertainties on trigger thresholds ($\simeq$~9\%),
variations of the input Monte Carlo distributions ($\simeq$~9\%)
and track reconstruction uncertainties especially at low $p_T$ ($\simeq$~6\%).
In particular for the $\rho^0$ analysis,
where the mass distribution is skewed compared to a Breit-Wigner shape,
uncertainties arise due to the assumptions of the fit for the interference 
between the resonant signal and the non-resonant background contributions
($\simeq$~7\%). 
Other significant contributions to the uncertainty are contamination due to 
$e$-gas interactions ($\simeq$~2-5\%) and from higher mass dissociated 
photon states, such as elastic $\omega$ and $\phi$ decays ($\simeq$~2-7\%).
The uncertainty due to neglecting radiative corrections can also be
estimated to be $\simeq$ 4-5\%~\cite{Hrho,Zrho}.

Finally, one of the key problems in obtaining accurate
measurements of the exclusive cross sections and the $t$ slopes 
is the uncertainty on the double dissociation component, where the 
proton has also dissociated into a low mass nucleon system~\cite{dd}. 
The forward calorimeters will see the dissociation
products of the proton if the invariant mass of the nucleon system, $M_N$,
is above approximately 4~GeV. 
A significant fraction of double dissociation events produce a limited mass 
system which is therefore not detected.
Proton remnant taggers are now being used further down 
the proton beamline to provide constraints on this fraction and, in
the H1 experiment, further constraints are provided by measuring 
secondary interactions in the forward muon system.
Based on $p\bar{p}$ data one finds that the dissociated mass spectrum falls 
as $ dN/dM_N^2 = 1/M_N^n $
with $n$ = 2.20 $\pm$ 0.03 at $\sqrt{s} = 1800$~GeV from CDF 
measurements~\cite{CDF2}.
However it should be noted that this measurement corresponds to a restricted 
mass interval. The extrapolation to lower masses is subject to
uncertainties and the universality of this dissociation is open to experimental
question, given the different behaviour at the upper vertex.
Precisely how the proton dissociates and whether the proton
can be regarded as dissociating independently of the photon system is not 
a priori known. Currently, this uncertainty is reflected in the cross sections
by allowing the value of $n$ to vary from around 2 to 3, although this
choice is somewhat arbitrary.
The magnitude of the total double dissociation contribution is estimated to 
be typically $\simeq 50\%$ prior to cuts on forward energy deposition, a 
value which can be cross-checked in the data with an overall uncertainty 
of $\simeq~10\%$ which is due to the considerations above.
Combining the above uncertainties, the overall systematic errors in the 
various cross sections are typically $\simeq~20\%$.

Photoproduction processes have been extensively studied in 
fixed-target experiments,
providing a large range in $W$ over which to study the cross sections.
The key features are the weak dependence of the cross section on $W$, 
an exponential dependence on $t$ with a slope which shrinks with increasing $W$
and the retention of the helicity of the photon by the vector meson. 
The $t$ dependence of the $\rho^0$ photoproduction data is illustrated in 
Figure~2 where the H1 and ZEUS data are 
compared to a compilation of lower energy data~\cite{aston}.
The data are consistent with a shrinkage
of the $t$ slope with increasing $E_\gamma \simeq W^2/2 $, where
$E_\gamma$ is the photon energy in the proton rest frame,
as indicated by the curve for soft pomeron exchange~\cite{schuler}.

\begin{figure}
\epsfxsize=8.cm
\centering
\leavevmode
\epsfbox[119 328 442 569]{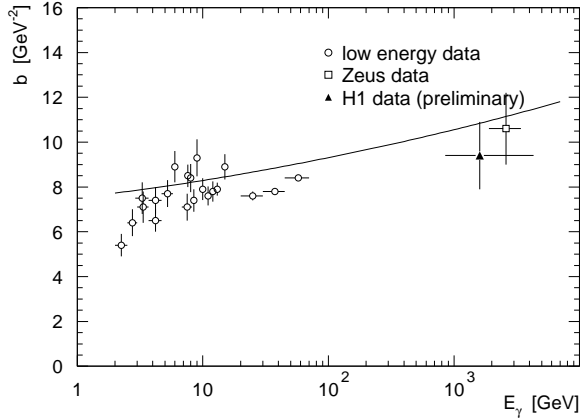}
\caption{Dependence of the exponential slope parameter $b$ as a function 
of $E_\gamma$ for exclusive $\rho^0$ photoproduction compared to the
soft pomeron exchange prediction of Schuler and Sj\"ostrand.}
\end{figure}

The measured $t$ slopes are 
$9.4\pm1.4\pm0.7$~GeV$^{-2}$~(H1)~\cite{Hrho} and 
$10.4\pm0.6\pm1.1$~GeV$^{-2}$~(ZEUS)~\cite{Zrho} for the 
$\rho^0$ (where similar single-exponential fits have been 
applied). These values can be compared to 
$7.3\pm1.0\pm0.8$~GeV$^{-2}$~(ZEUS)~\cite{Zphi} for the $\phi$ and 
$4.7\pm1.9$~GeV$^{-2}$~(H1)~\cite{Hpsi} 
for the $J/\psi$.
Physically, the slope of the $t$ dependence in diffractive interactions 
tells us about the effective
radius of that interaction, $R$: 
if d$\sigma/dt \propto e^{-b|t|}$, then
b $\simeq$ 1/4 $R^2$. The range of measured $b$ slopes varies from 
around 4~GeV$^{-2}$ ($R \simeq 0.8$~fm) to 10~GeV$^{-2}$ ($R \simeq 1.3$~fm). 
Further, the interaction radius can be approximately related to the
radii of the interacting proton and vector meson, 
$R \simeq \sqrt{R_P^2 + R_V^2}$.
Given $R_P \simeq 0.7$~fm, 
then this variation in $b$ slopes corresponds to a significant change in 
the effective radius of the interacting vector meson from 
$R_V \simeq 0.4$~fm to $R_V \simeq 1.1$~fm. 

\begin{figure}[htb]
\epsfxsize=10.cm
\centering
\leavevmode
\epsfbox{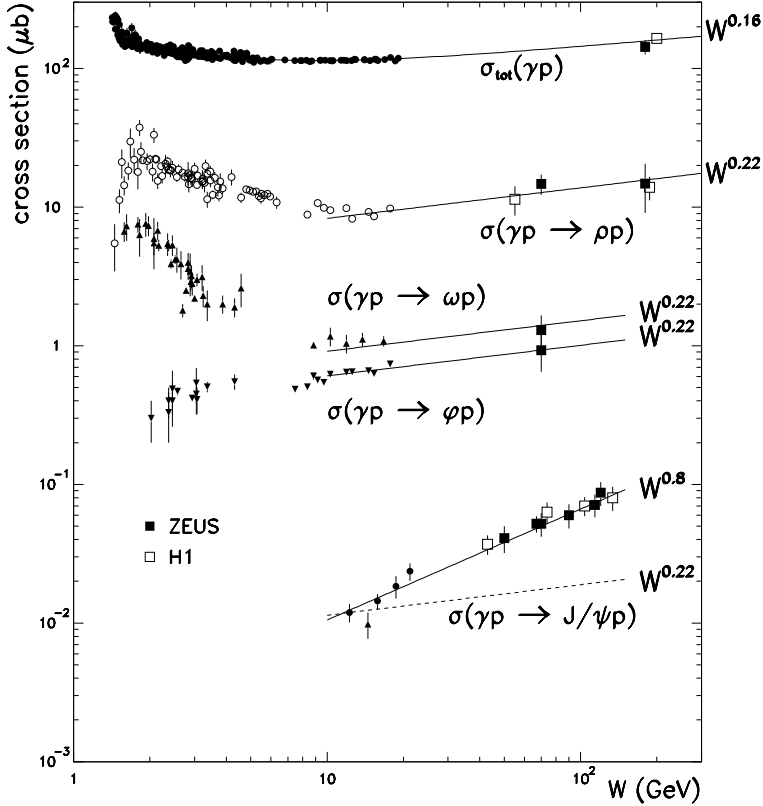}
\caption{$W$ dependence of the exclusive vector meson and total 
photoproduction cross sections compared to various power law dependences
discussed in the text.}
\end{figure}         

Integrating over the measured $t$ dependence, 
the $W$ dependence of the results on 
exclusive vector meson photoproduction cross sections 
are shown in Figure~3~\cite{levy}. From the experimental perspective,
there is generally good agreement on the measured cross sections.
The $\gamma p$ total cross section is also shown in Figure~3, 
rising with increasing 
energy as in hadron-hadron collisions and consistent with a value of 
$\epsilon \simeq 0.08$ i.e. the total cross section increases as $W^{0.16}$. 

Given the dominance of the pomeron trajectory at high $W$ and
an approximately exponential behaviour of the $|t|$ distribution with slope 
$b \simeq 10$, whose mean $|\bar{t}|$ value
is given by $1/b$, the diffractive cross section
rise is moderated from
$$W^{4\epsilon}= W^{0.32}$$
to 
$$W^{4(\epsilon - \alpha'|\bar{t}|)} \equiv W^{4\bar{\epsilon}} = W^{0.22}.$$
Here $\bar{\epsilon} = 0.055$ characterises the effective energy dependence 
after integration over $t$. 
The observed shrinkage of the diffractive peak 
therefore corresponds to a relative reduction
of the diffractive cross section with increasing energy. 
Such a dependence describes the general increase of the 
$\rho^0$, $\omega$ and $\phi$ vector meson cross sections with increasing $W$.
However, the rise of the $J/\psi$ cross section is clearly not described 
by such a $W$ dependence, the increase being described by an 
effective $W^{0.8}$ dependence. Whilst these effective powers are for
illustrative purposes only, it is clear that in exclusive $J/\psi$ production
a new phenomenon is occurring.

Qualitatively, the $W^{0.8}$ dependence, corresponding to 
$\bar{\epsilon} \simeq 0.2$, 
could be ascribed 
to the rise of the gluon density observed in the scaling violations of
$F_2$. 
The $J/\psi$ mass scale, $M^2$, is larger than the QCD scale 
$\Lambda^2$, and it is therefore possible to apply pQCD 
techniques.
Quantitatively, the theoretical analysis predicts that 
the rise of the cross section is proportional to the square of the gluon
density at small-$x$ and allows discrimination among the latest 
parametrisations
of the proton structure function~\cite{ryskin}.
We also know from measurements of the DIS $\gamma^* p$ total cross 
section that application of formula (1) results in a value of $\epsilon$ 
which increases with increasing $Q^2$, with $\epsilon \simeq$ 0.2 to 0.25 
at $Q^2\simeq10$~GeV$^2$~\cite{levy}.
The fact that the corresponding relative rise of $F_2$ with decreasing 
$x$ can be described by pQCD evolution~\cite{GRV}
points towards a calculable function $\epsilon = \epsilon(Q^2)$
for $Q^2 \sgeq Q_o^2 \simeq 0.3~$GeV$^2$.

One contribution to the DIS $\gamma^* p$ total cross section is
the electroproduction of low mass vector mesons.
Experimentally, the statistical errors typically dominate with 
systematic uncertainties similar to the
photoproduction case. The trigger uncertainties are significantly
reduced, however, since the scattered electron is easily identified
and the radiative
corrections, which are more significant ($\simeq 15\%$~\cite{wulff}),
can be corrected for. 
The $W$ dependence of the DIS $\rho^0$ and $\phi$ cross sections
for finite values of $Q^2$ are shown in Figure~4,
compared to the corresponding photoproduction cross sections.
The $W$ dependence for the $\rho^0$ and $\phi$ electroproduction 
data are similar to those for
the $J/\psi$ photoproduction data, consistent
with an approximate $W^{0.8}$ dependence also shown in Figure~4.
An important point to emphasise here is that the relative production of 
$\phi$ to $\rho^0$ mesons approaches the quark model prediction of 2/9 at large
$W$ and large $Q^2$, which would indicate the applicability 
of pQCD to these cross sections.
The measurements
of the helicity angle of the vector meson decay provide a measurement of 
$R= \sigma_L/\sigma_T$ for the (virtual) photon, assuming s-channel helicity
conservation, i.e. that the vector meson preserves the helicity of the photon.
The photoproduction measurements for the $\rho^0$ are consistent with the 
interaction of 
dominantly transversely polarised photons
($R=0.06\pm0.03$~(ZEUS)~\cite{Zrho}). However, adopting the same 
analysis for virtual photons, 
$R=1.5^{+2.8}_{-0.6}$~(ZEUS)~\cite{Zrho*}, 
inconsistent with the behaviour in photoproduction and
consistent with a predominantly longitudinal exchange. 
This predominance is expected for an underlying 
interaction of the virtual photon with the constituent quarks of the $\rho^0$.
Also, the measured 
$b$ slope approximately halves
from the photoproduction case to a value of
$b=5.1^{+1.2}_{-0.9}\pm1.0$~(ZEUS)~\cite{Zrho*}, comparable to that
in the photoproduced $J/\psi$ case.
The
basic interaction is probing smaller distances, which allows a first comparison
of the observed cross section with the predictions of leading-log 
pQCD~(see~\cite{Zrho*}).

\begin{figure}
\begin{minipage}[htb]{8cm}
\centering
~~(a)
\end{minipage}
\begin{minipage}[htb]{8cm}
\centering
~~~(b)
\end{minipage}
\\
\begin{minipage}[htb]{8cm}
\epsfxsize=7cm
\centering
\leavevmode
\epsfbox{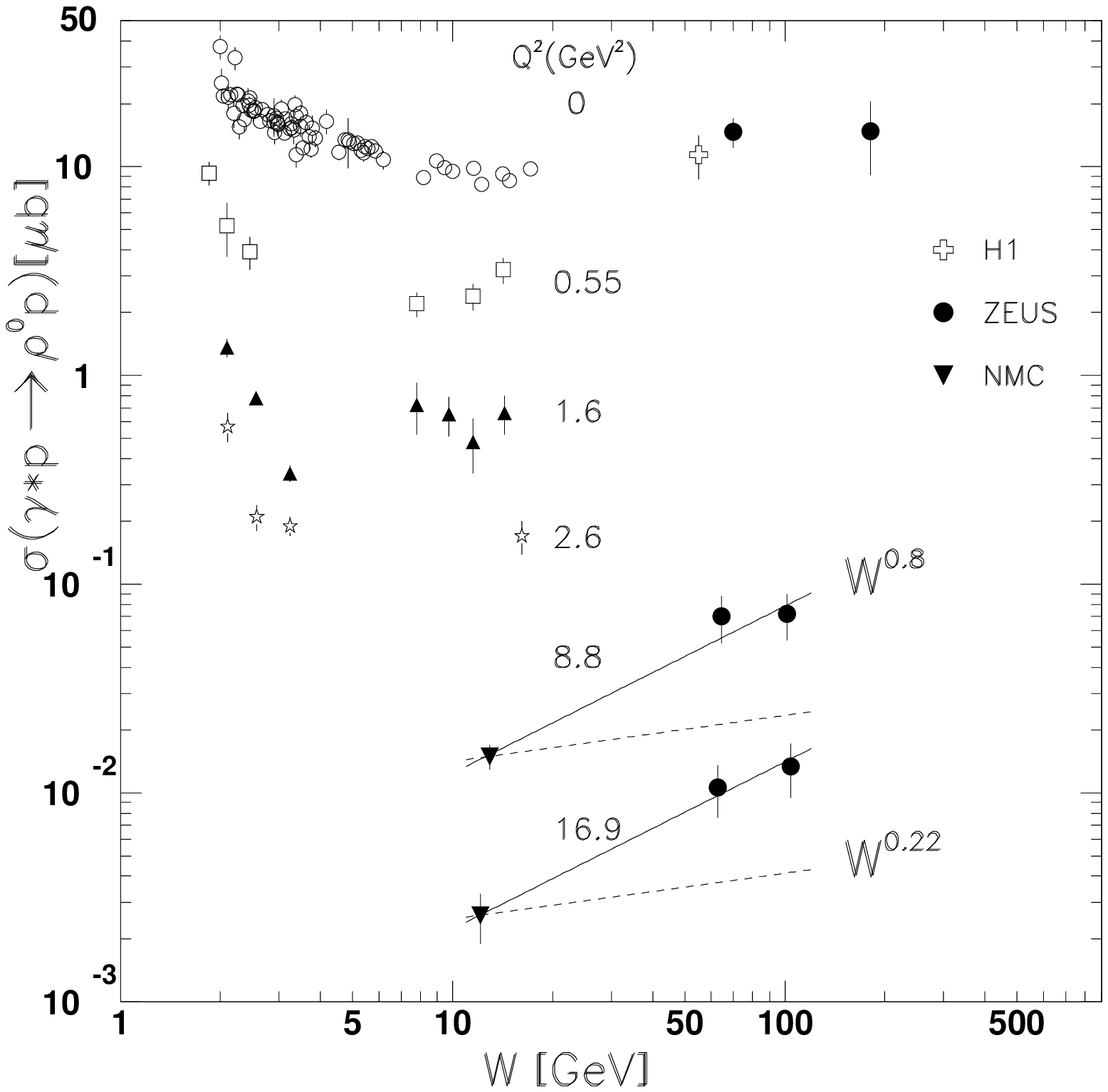}
\end{minipage}
\begin{minipage}[htb]{8cm}
\epsfxsize=7cm
\centering
\leavevmode
\epsfbox{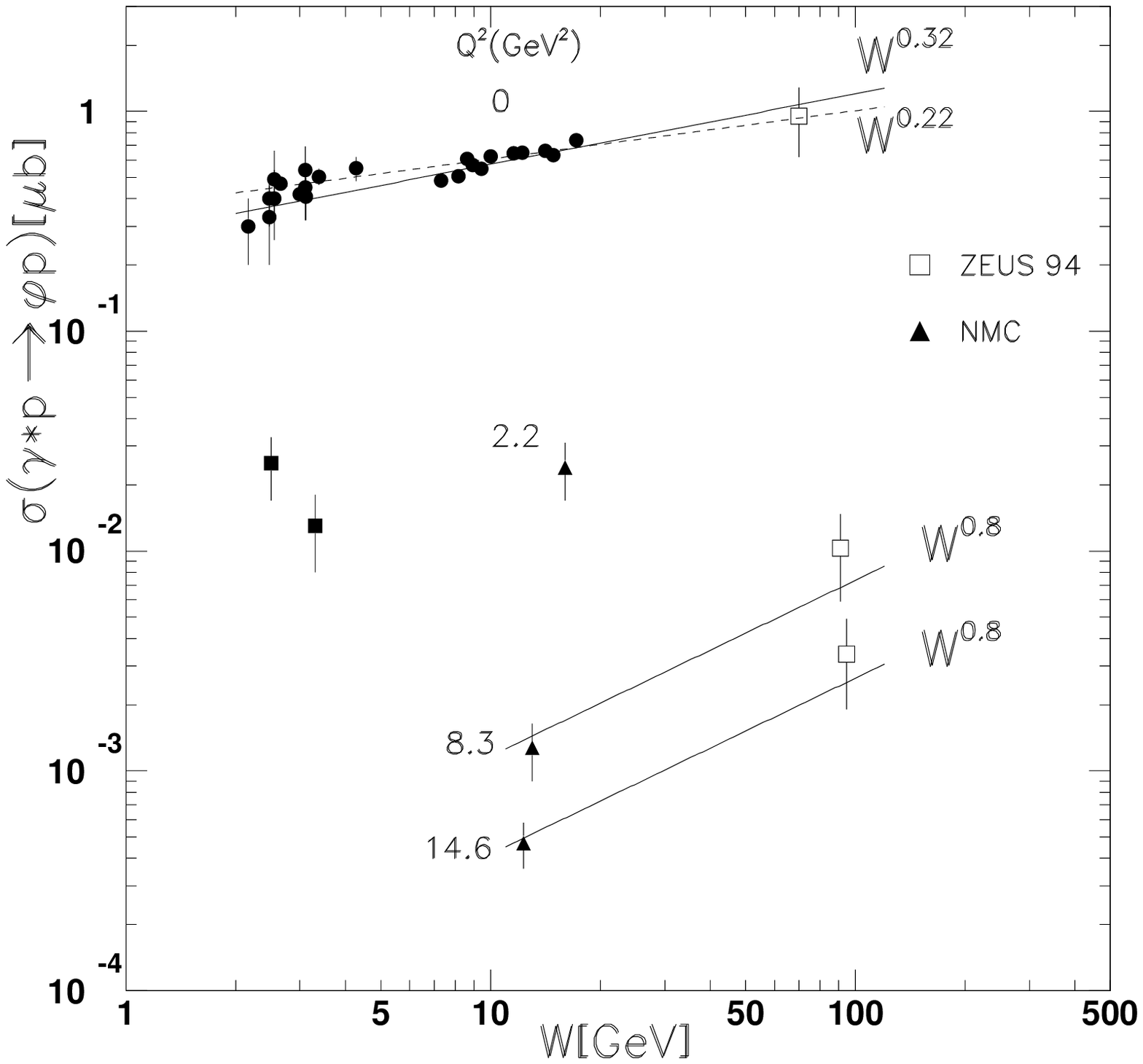}
\end{minipage}
\\
\caption{$W$ dependence of exclusive (a) $\rho^0$ and (b) $\phi$ 
electroproduction cross sections for fixed values of $Q^2$ compared
to various power law dependences discussed in the text.}
\end{figure}

Finally, first results based on the observation of 42 $J/\psi$ events at
significant $<\!Q^2\!>=17.7$~GeV$^2$ have been reported by H1~\cite{Hpsi*}.
The cross section has been evaluated in two $W$ intervals in order 
to obtain an indication of the $W$ dependence, as shown in
Figure~5, where an estimated 50\% contribution due to double dissociation has
been subtracted~\cite{berger}.
The electroproduction data are shown with statistical errors only
although the systematics are estimated to be smaller than these errors
($\simeq 20\%$). 
The electroproduction and photoproduction $J/\psi$ data are consistent with 
the $W^{0.8}$ dependence ($\bar{\epsilon} \simeq 0.2$) noted previously.
The $J/\psi$ electroproduction cross section is of the same order of 
that of the $\rho^0$ data, in marked contrast to the significantly lower 
photoproduction cross section for the $J/\psi$, even at HERA energies, 
also shown in Figure~5.
Further results in this area would allow tests of the underlying dynamics 
for transverse and longitudinally polarised photons coupling to light
and heavy quarks in the pQCD calculations.

\begin{figure}
\epsfxsize=7.cm
\centering
\leavevmode
\epsfbox[13 53 437 472]{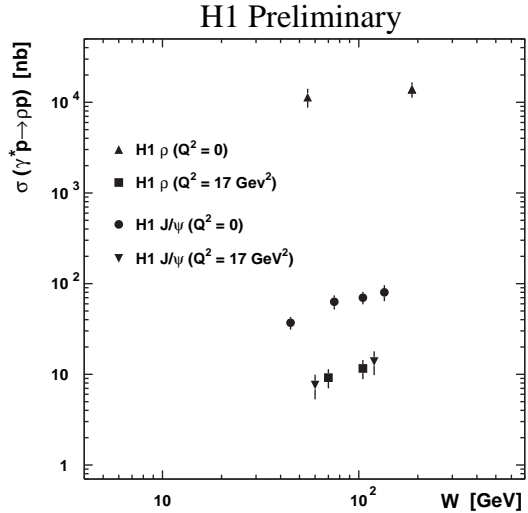}
\caption{H1 measurements of the $W$ dependence of 
electroproduction and photoproduction cross sections of exclusive vector 
mesons.}
\end{figure}

In conclusion, there is an accumulating body of exclusive vector meson
production data, measured with a systematic precision of $\simeq 20\%$, which
exhibit two classes of $W^2$ behaviour: a slow rise consistent with that
of previously measured diffractive data for low $M^2$ photoproduction
data but a significant rise of these cross sections 
when a finite $Q^2$ and/or a significant $M^2$ is measured.

\subsection{Deep Inelastic Structure of Diffraction}
One of the major advances in the subject of diffraction has
been the observation of large rapidity gap events in DIS and their 
subsequent analysis in terms of a diffractive structure 
function~\cite{Hd1,Zd1}. In these analyses,
the signature of diffraction is the rapidity gap,
defined by measuring the maximum pseudorapidity of the 
most-forward going particle
with energy above 400~MeV, $\eta_{max}$, 
and requiring this to be well away from
the outgoing proton direction. A typical requirement of 
$\eta_{max} < 1.5$ corresponds to a low 
mass state measured in the detectors of $\ln(M^2) \sim 3$ units 
and a large gap of 
$\ln(W^2) - \ln(M^2) \sim 7$ units with respect to the 
outgoing proton (nucleon system).
In order to increase the lever arm in $M^2$, 
the H1 and ZEUS analyses have extended the $\eta_{max}$ cuts to 3.2 
and 2.5, respectively.
This is achieved directly using the forward muon system/proton 
remnant taggers, in the case of H1, 
or via the measurement of a further discriminating 
variable, $\cos\theta_H=\sum_i p_{z_i}/|\sum_i \vec{p_i}|$,
where $\vec{p_i}$ is the momentum vector of a calorimeter cell,
for ZEUS.
These extensions are, however, at the expense of a 
significant non-diffractive DIS background 
(up to $\simeq 50\%$ and $\simeq 20\%$, respectively).
In each case, this background is estimated using the 
the colour-dipole model as implemented in the 
ARIADNE~4.03 program~\cite{ariadne}, which 
reasonably reproduces the observed forward $E_T$ flows in
non-diffractive interactions. 
The uncertainty on this background is estimated by changing the applied
cuts or by using other Monte Carlo models and is up to 20\% for large
masses, $M^2$, of the dissociated photon.
The double dissociation contribution is estimated with 
similar uncertainties to the vector meson case.
Other systematic errors 
are similar to those for the $F_2$ analyses ($\sleq 10\%$) 
with additional acceptance uncertainties due to
variations of the input diffractive Monte Carlo distributions.

In the presentation of the results, the formalism changes~\cite{ingprytz},
reflecting an assumed underlying partonic description,
and two orthogonal variables are determined: 
$$ \xpom = \frac{(P-P')\cdot q}{P\cdot q} 
\simeq \frac{M^2 + Q^2}{W^2 + Q^2}~~~~~~~~~
\beta = \frac{Q^2}{2(P-P')\cdot q} \simeq \frac{Q^2}{M^2 + Q^2},$$
the momentum fraction of the pomeron within the proton and  
the momentum fraction of the struck quark within the pomeron,
respectively. 
The structure function is then defined by analogy to that of the 
total $ep$ cross section:
$$
  \frac{d^3\sigma_{diff}}{d\beta dQ^2 d\xpom} = \frac{2 \pi
    \alpha^2}{\beta Q^4} \; (1+(1-y)^2) \;
  F_2^{D(3)}(\beta,Q^2,\xpom),
$$
where 
the contribution of $F_L$ and radiative corrections are neglected
and an integration over the (unmeasured) $t$ variable is performed.
The effect of neglecting $F_L$ corresponds to a relative reduction of the 
cross section at small \xpom~(high $W^2$) which is always $<17\%$ and therefore
smaller than the typical measurement uncertainties ($\simeq~20\%$).

As discussed above, a major uncertainty comes from the estimation of the 
non-diffractive background. This problem has been addressed in a different
way in a further analysis by ZEUS~\cite{Zd2}.
In this analysis the mass spectrum, $M^2$, 
is measured as a function of $W$ and $Q^2$, 
as shown in Figure~6 for four representative intervals, where
the measured mass is reconstructed in the calorimeter and 
corrected for energy loss but not for 
detector acceptance, resulting in the turnover at large $M^2$. 
The diffractive data are observed as a low mass shoulder at low $W$,
which becomes increasingly apparent at higher $W$.
Also shown in the figure are the estimates of the non-diffractive background
based on (a) the ARIADNE Monte Carlo (dotted histogram) and (b) a direct
fit to the data, discussed below.

\begin{figure}
\epsfxsize=8.cm
\centering
\leavevmode
\epsfbox[17 154 532 674]{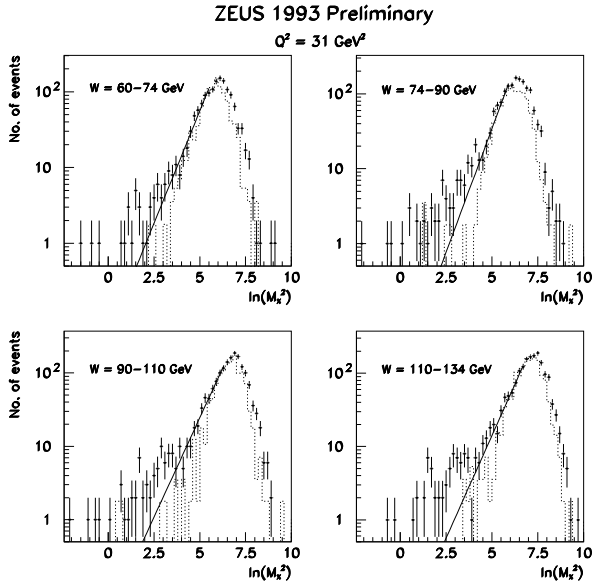}
\caption{Preliminary ZEUS analysis of the $\ln M^2$ distributions as
a function of $W$ at $Q^2 = 31$~GeV$^2$. 
The solid lines show the extrapolation 
of the nondiffractive background as determined by the fits discussed in
the text. 
The dotted histograms show the predictions for non-diffractive scattering as
modelled using the ARIADNE~4.03 program.}
\end{figure}

The probability of producing a gap is exponentially suppressed 
as a function of the rapidity gap, and hence as a function of $\ln(M^2$),
for non-diffractive interactions. The slope of this exponential
is directly related to the height of the plateau distribution of 
multiplicity in the region of rapidity where the subtraction is made.
The data can thus be fitted to functions of the form
$ dN/d \ln(M^2) = D + C {\rm exp}( b \cdot \ln(M^2)) $, 
in the region where the detector acceptance is uniform,
where $b$, $C$ and $D$ are determined from the fits.
Here, $D$ represents a first-order estimate of the diffractive contribution
which is flat in $\ln(M^2$). The important parameter is $b$, which is
determined to be $b = 1.55\pm0.15$ in fits to each of the measured data 
intervals, compared to $b=1.9\pm 0.1$ estimated from the ARIADNE Monte Carlo.
The systematic uncertainty in the background reflects various changes to the 
fits, but in each case the measured slope is incompatible with that of the 
Monte Carlo.
This result in itself is interesting, since the fact that ARIADNE approximately
reproduces the observed forward $E_T$ ($\sim$ multiplicity) flow but does not
reproduce the measured $b$ slope suggests that significantly different 
correlations of the multiplicities are present in non-diffractive DIS 
compared to the Monte Carlo expectations. 
Also new in this analysis is that
the diffractive Monte Carlo POMPYT~1.0~\cite{pompyt} has been tuned
to the observed data contribution for low mass states, allowing the high
$\beta$ region to be measured up to the kinematic limit ($\beta\rightarrow 1$)
and radiative corrections have been estimated in each interval 
($\sleq 10\%$~\cite{wulff}). 

The virtual-photon proton cross sections 
measured at fixed $M^2$ and $W$, measured in this analysis,
can be converted to $F_2^{D(3)}$ at fixed $\beta$ and $\xpom$.
These results are shown in Figure~7 as the ZEUS(BGD)~\cite{Zd2} analysis,
compared to the earlier
H1~\cite{Hd1} and ZEUS(BGMC)~\cite{Zd1} analyses in comparable intervals
of $\beta$ and $Q^2$ as a function of $\xpom$. The overall 
cross sections in each $\beta$ and $Q^2$ interval are similar, however,
the \xpom~dependences are different.
As can be seen in Figure~6, the background estimates are significantly
different which results in a systematic shift in 
the $W$ (\xpom) dependence at fixed $M$ ($\beta$) and $Q^2$.

\begin{figure}
\epsfxsize=10.cm
\centering
\leavevmode
\epsfbox[6 111 528 694]{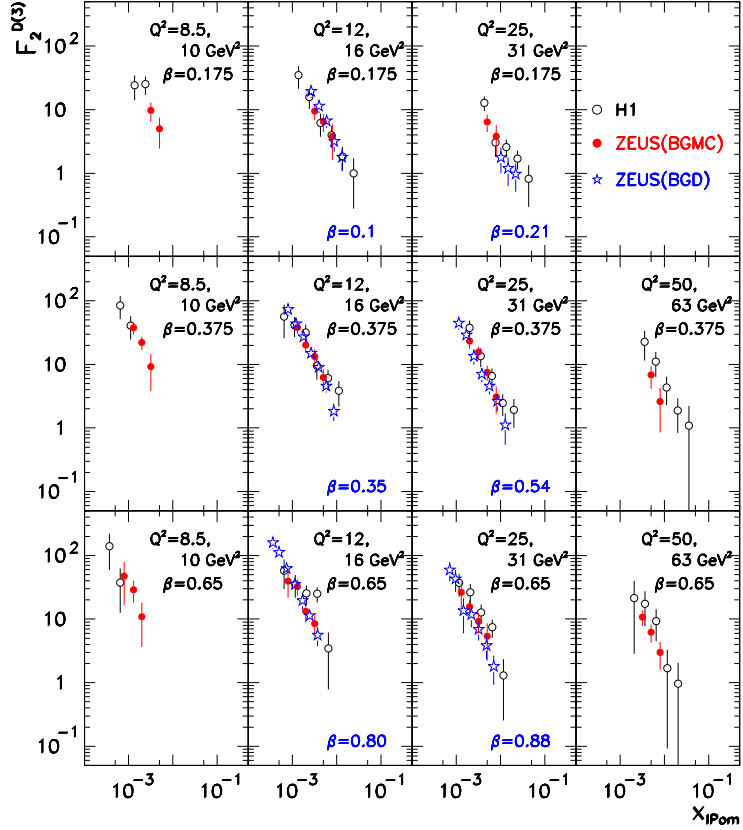}
\caption{Comparison of the HERA data for $F_2^{D(3)}$ as function of
\xpom\ for the H1 and ZEUS(BGMC) analyses where the Monte Carlos
are used to estimate the background. The upper (lower) $Q^2$ value refers to 
the H1 (ZEUS) analysis. The preliminary ZEUS(BGD) where a fit to the data
is used to estimate the non-diffractive background is compared at slightly
different $\beta$ values noted at the bottom of the figure.}
\end{figure}

Fits of the form 
$F_2^{D(3)} = b_i \cdot \xpom^{n}$
are performed 
where the normalisation constants $b_i$ are allowed to differ
in each $\beta,Q^2$ interval. 
The fits are motivated by the factorisable ansatz of
$F_2^{D(3)}(\beta, Q^2, \xpom) 
= f_{\pom}(\xpom) \cdot F_2^{\pom}(\beta,Q^2), $
where $f_{\pom}(\xpom)$ measures the flux of pomerons in the proton
and $F_2^{\pom}(\beta,Q^2)$ is the probed structure of the pomeron.
The exponent of \xpom~is identified as $n = 1+2\cdot\bar{\epsilon}$,
where $\bar{\epsilon}$ measures the effective \xpom~dependence 
($\equiv W^2$ dependence at fixed $M^2$ and $Q^2$) of the cross section, 
integrated over $t$, as discussed in relation to exclusive vector meson
production. 
In each case, the $\chi^2/DOF$ are $\simeq 1$ indicating that a single 
power law dependence on energy provides a reasonable description of the data 
and that effects due to factorisation breaking 
predicted in QCD-based calculations~\cite{nz} are not yet observable.
The results for $\bar{\epsilon}$ are
$0.095\pm0.030\pm0.035$~(H1)~\cite{Hd1},
$0.15\pm0.04^{+0.04}_{-0.07}$~(ZEUS(BGMC))~\cite{Zd1} and
$0.24\pm0.02^{+0.07}_{-0.05}$~(ZEUS(BGD))~\cite{Zd2},
where the systematic errors are obtained by refitting according to a series
of systematic checks outlined above.
It should be noted that the (2$\sigma$) systematic shift between the 
ZEUS(BGD) and ZEUS(BGMC) can be attributed to the method of background 
subtraction.
Whilst the H1 and ZEUS(BGMC) analyses, based on Monte Carlo background 
subtraction, agree within errors, the ZEUS(BGD) value 
is different from the H1 value at the 3$\sigma$ level.

The Donnachie-Landshoff prediction~\cite{dl} is $\bar{\epsilon} 
\simeq 0.05$, after integration over an assumed $t$ dependence and taking 
into account shrinkage. 
While comparison with the H1 value indicates that this
contribution is significant, the possibility of additional
contributions cannot be neglected.
Taking the ZEUS(BGD) value, this measurement is incompatible 
with the predicted soft pomeron behaviour at the 4$\sigma$ level.
Estimates of the effect of $\sigma_L$ made by 
assuming $\sigma_L = (Q^2/M^2) \sigma_T$ rather than $\sigma_L = 0$
result in $\bar{\epsilon}$ increasing from 0.24 to 0.29. 

The values can also be compared with $\bar{\epsilon} \simeq$ 0.2
obtained from the exclusive photoproduction of $J/\psi$ mesons and the 
electroproduction data or with $\epsilon \simeq$ 0.2 to 0.25
obtained from the dependence of the total cross sections 
in the measured $Q^2$ range~\cite{levy}.
In the model of Buchm\"uller and Hebecker~\cite{buch}, 
the effective exchange is 
dominated by one of the two gluons. In terms
of $\epsilon$, where the optical theorem is no longer relevant,
the diffractive cross section
would therefore rise with an effective 
power which is halved to $\epsilon \simeq$ 0.1 
to 0.125.
The measured values are within the range of these estimates.
 
The overall cross sections in each $\beta$, $Q^2$ interval are similar
and one can integrate over the measured \xpom~dependence
in order to determine $\tilde{F}_2^D$($\beta, Q^2$), a quantity which measures
the internal structure of the pomeron up to an arbitrary integration 
constant. Presented in this integrated form, the data agree on the general 
features of the internal structure. In Figure~8 the H1 data are
compared to preliminary QCD fits~\cite{Hd2}. The general
conclusions from the $\beta$ dependence are that the pomeron has a 
predominantly hard structure, typically characterised by a 
symmetric $\beta(1-\beta)$ dependence, but also containing an additional,
significant contribution at low $\beta$ which has been fitted in the ZEUS
analysis~\cite{Zd1}. 
The virtual photon only couples directly to quarks, but the overall cross 
section can give indications only of the relative proportion of quarks and 
gluons within the pomeron, since the flux normalisation 
is somewhat arbitrary~\cite{Zd1}.
The $Q^2$ behaviour is broadly scaling, consistent with a partonic
structure of the pomeron. Probing more deeply, however, a characteristic 
logarithmic rise of $\tilde{F}_2^D$ is observed in all $\beta$ intervals.
Most significantly, at large $\beta$ a predominantly quark-like object
would radiate gluons resulting in negative scaling violations as in the 
case of the large-$x$ ($\sgeq 0.15$) behaviour of the proton.
The question of whether the pomeron is predominantly quarks or gluons,
corresponding to a ``quarkball" or a ``gluemoron"~\cite{cf},
has been tested quantitatively by H1 using QCD fits to 
$\tilde{F}_2^D$~\cite{Hd2}. A flavour singlet 
quark density input of the form $zq(z) = A_q \cdot z^{B_q}(1-z)^{C_q}$, where
$z$ is the momentum fraction carried by the quark, yields a numerically
acceptable $\chi^2$. The characteristic $Q^2$ behaviour, however, is not
reproduced. Adding a gluon contribution of similar form gives an excellent 
description of the data. The fit shown uses $B_q = 0.35$, $C_q = 0.35$,
$B_g = 8$, $C_g = 0.3$. In general, the fits tend to favour inputs where 
the gluon carries a significant fraction, $\sim$ 70 to 90\%,
of the pomeron's momentum.

\begin{figure}[htb]
\epsfxsize=10.cm
\centering
\leavevmode
\epsfbox[32 137 530 654]{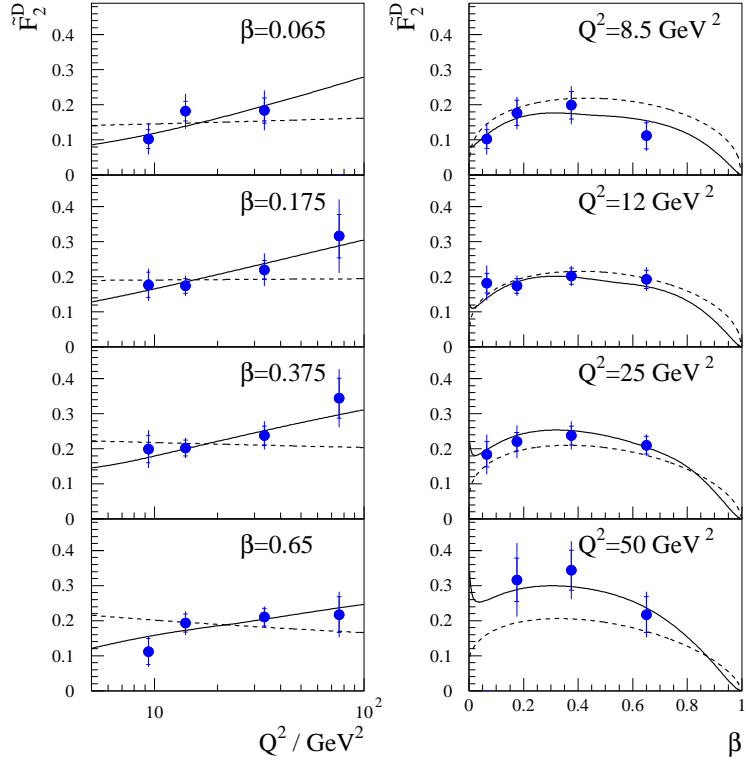}
\caption{H1 data on $\tilde{F}_2^D$($\beta, Q^2$) as a function
of $Q^2$ ($\beta$) at fixed $\beta$ ($Q^2$). The data are compared to
preliminary
leading-order QCD fits where:
(a) only quarks are considered at the starting scale,
$Q_0^2 = 4$~GeV$^2$, indicated by the dashed line 
($\chi^2/DOF = 13/12$, 37\% CL); (b) gluons also contribute at the
starting scale, resulting in a fit where gluons carry $\sim$ 90\%
of the momentum, indicated by the full line ($\chi^2/DOF = 4/9$, 91\% CL).
}
\end{figure}

\subsection{Jet structure}
The question of the constituent content of the pomeron can also be addressed
via measurements of diffractively produced jets in the photoproduction
data~\cite{Zjet}. Jets are reconstructed at large $W$ ($130< W < 270$~GeV) 
using the cone algorithm with cone radius 1 and $E_T^{jet} > 8$~GeV.
The diffractive
contribution is identified as a tail in the $\eta_{max}$ distribution 
of these events above the 
PYTHIA~5.7~\cite{pythia} Monte Carlo expectation. In Figure 9(a) the measured
cross section is compared to various model predictions as a function of the 
jet rapidity. 
Comparison with the non-diffractive contribution estimated from
PYTHIA indicates a significant excess at lower values of $\eta^{jet}$.
Here, standard photon and proton parton distributions are adopted and
the overall scale, which agrees with the non-diffractive data normalisation,
is set by $E_T^{jet}$. Also shown are the predicted diffractive cross sections
from POMPYT using a hard ($z(1-z)$) quark, hard gluon or soft
(($1-z)^5$) gluon where a Donnachie-Landshoff flux factor is adopted
and the momentum sum rule is assumed to be satisfied in each case. 
Sampling low-energy (soft) gluons corresponds to a small cross section
and can be discounted, 
whereas high-energy (hard) gluons and/or quarks can account for the cross 
section by changing the relative weights of each contribution.
The $x_\gamma$ distribution for these events, where $x_\gamma$ is the 
reconstructed momentum fraction of the interacting photon, 
is peaked around 1, indicating that at these $E_T^{jet}$ values
a significant fraction of events is due to direct processes where the 
whole photon probes the pomeron constituents.

We now have two sets of data, the DIS data~\cite{Zd1} probing the 
pomeron structure
at a scale $Q$ and the jet data probing at a scale 
of $E_T^{jet}$. Each probes
the large $z$ structure of the pomeron with 
the jet and DIS data, 
predominantly sampling the (hard) gluon and quark distributions, respectively.
In Figure~9(b) the preferred momentum 
fraction carried by the (hard) gluon, $c_g$, is 
indicated by the overlapping region of the jet (dark band) and DIS 
(light band) fits to the data. 
Considering the systematics due to the non-diffractive background, modelled
using the Monte Carlo models, a range of values consistent with 
$c_g \sim 0.55 \pm 0.25$ can be estimated.
The result depends on the assumption that the cross sections for both sets
of data factorise with 
a universal flux, characterised by the same value of $\bar{\epsilon}$
in this $W$ range, 
but does not assume the momentum sum rule.

\begin{figure}
\begin{minipage}[htb]{8cm}
\centering
~~~~~~~(a)
\end{minipage}
\begin{minipage}[htb]{8cm}
\centering
~~~~~~~(b)
\end{minipage}
\\
\begin{minipage}[htb]{8cm}
\epsfxsize=7cm
\centering
\leavevmode
\epsfbox{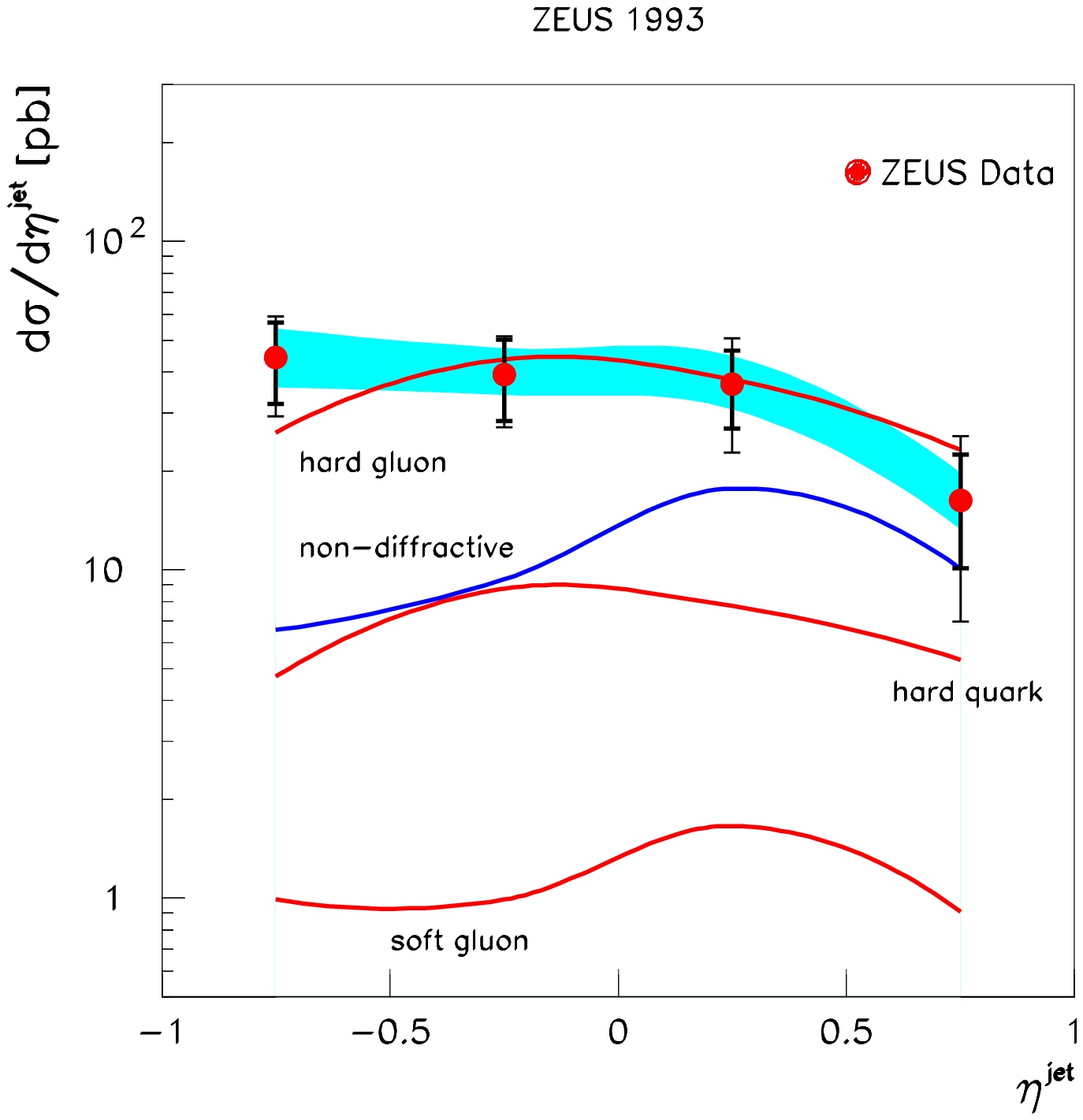}
\end{minipage}
\begin{minipage}[htb]{8cm}
\epsfxsize=7cm
\centering
\leavevmode
\epsfbox{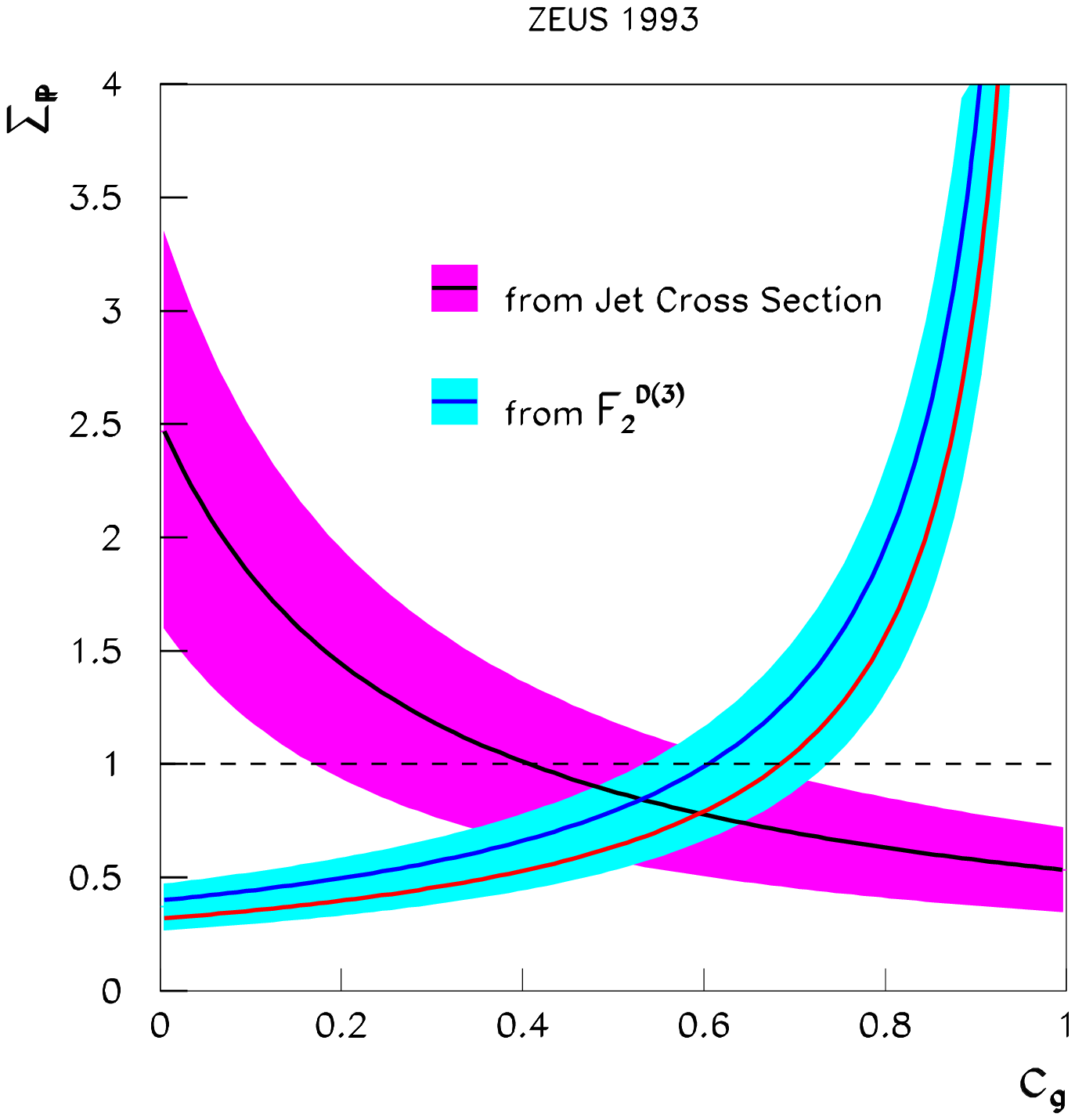}
\end{minipage}
\\
\caption{(a) Jet cross section as a function of jet rapidity for
events with $\eta_{max} < 1.8$. (b) Momentum sum rule assuming a
Donnachie-Landshoff flux, $\Spom$, versus the momentum fraction carried
by the gluons in the pomeron, $c_g$. The dark (light) error bands correspond to 
statistical errors on the fits to the jet (DIS) data discussed in the text.}
\end{figure}

So far we have only considered the case of small-$t$ diffraction with respect
to the outgoing proton. Further insight into the diffractive exchange process
can be obtained by measurements of the rapidity gap between jets. Here, 
a class of events is observed with little hadronic
activity between the jets~\cite{Zt}. 
The jets have $E_T^{jet} > 6$~GeV and are separated by a pseudorapidity 
interval ($\Delta\eta$) of up to 4 units.
The scale of the momentum transfer, $t$, is not precisely defined but 
is of order $(E_T^{jet})^2$.
A gap is defined as the absence of particles with
transverse energy greater than 300~MeV between the jets.
The fraction of events containing a gap is then measured as a function of
$\Delta\eta$, as shown in Figure~10.  
The fit indicates the sum of an exponential behaviour, as
expected for non-diffractive processes and discussed in relation to the 
diffractive DIS data, and a flat distribution expected for diffractive
processes. At 
values of $\Delta\eta \sgeq 3$, an excess is seen with a constant fraction
over the expectation for non-diffractive exchange 
at $\simeq 0.07\pm 0.03$.
This can be interpreted as evidence for large-$t$ diffractive scattering.
In fact, secondary interactions of the photon and proton remnant
jets could fill in the gap and therefore the underlying process could play
a more significant r\^ole. 
The size of this fraction is relatively large when compared to a similar
analysis by D0 and CDF where a constant fraction at $\simeq 
0.01$ is observed~\cite{D0,CDF3}.
The relative probability may differ 
due to the higher $W$ values of the Tevatron compared to HERA or, perhaps, 
due to differences in the underlying $\gamma p$ and $p\bar{p}$ interactions.

\begin{figure}
\epsfxsize=6.cm
\centering
\leavevmode
\epsfbox{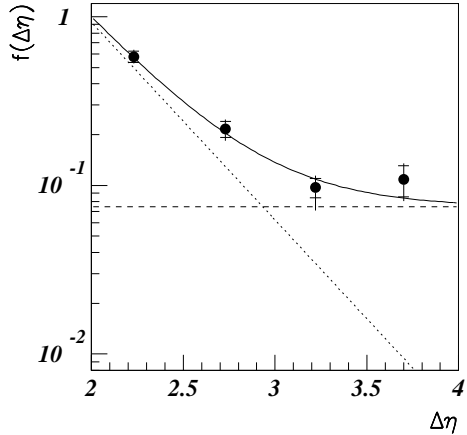}
\caption{Gap-fraction, $f(\Delta\eta)$, as a function of the rapidity gap
between the two jets compared with the result of a fit to an exponential plus 
a constant.}
\end{figure}

\subsection{Leading proton spectrometer measurements}
The advent of the leading proton spectrometers (LPS) at HERA is especially
important in these diffractive measurements, since internal
cross checks of the measurements as a function of $t$, $M^2$, $W^2$ and $Q^2$
can ultimately be performed 
and underlying assumptions can be questioned experimentally. Only in these
measurements can we positively identify the diffracted proton and hence 
substantially reduce uncertainties on the non-diffractive and double 
dissociation backgrounds. However, new uncertainties are introduced
due to the need for precise
understanding of the beam optics and relative alignment of the detectors.
Reduced statistical precision also results due to the 
geometrical acceptance of the detectors ($\simeq$~6\%).
In figure~11(a), various observed distributions are shown for 
240 events selected from the DIS data, with $Q^2 \sgeq 4$~GeV$^2$. 
The momentum distribution clearly indicates a significant diffractive 
peak at $E_p = 820$~GeV above the non-diffractive background and the observed
$M$ and $W$ distributions are well described by the NZ Monte Carlo~\cite{ada}.
The distribution of $\beta$ versus $\xpom$ indicates a significant fraction
of events at small $\beta \sleq 0.1$ which are difficult to access using the 
experimental techniques described earlier. The measurements are currently
being analysed, but a preliminary result on the $t$-dependence is shown in 
Figure~11(b), measured in a relatively high observed 
mass interval, $<\!M\!> = 9$~GeV,
at relatively low $Q^2$. The slope can be characterised by 
a single exponential fit with 
$b = 7.52 \pm 0.95^{+0.65}_{-0.82}$. 
This is somewhat high compared to 
the value of $b\simeq 4.5$ expected for a predominantly hard pomeron
but lies within the range of expectations of $4 \sleq b \sleq 10$.
However, before drawing conclusions, we should perhaps wait for further 
results on the general dependences measured in the LPS.

\begin{figure}
\begin{minipage}[htb]{9cm}
\centering
(a)
\end{minipage}
\begin{minipage}[htb]{7cm}
\centering
(b)
\end{minipage}
\\
\begin{minipage}[htb]{9cm}
\epsfxsize=8cm
\centering
\leavevmode
\epsfbox[19 191 539 697]{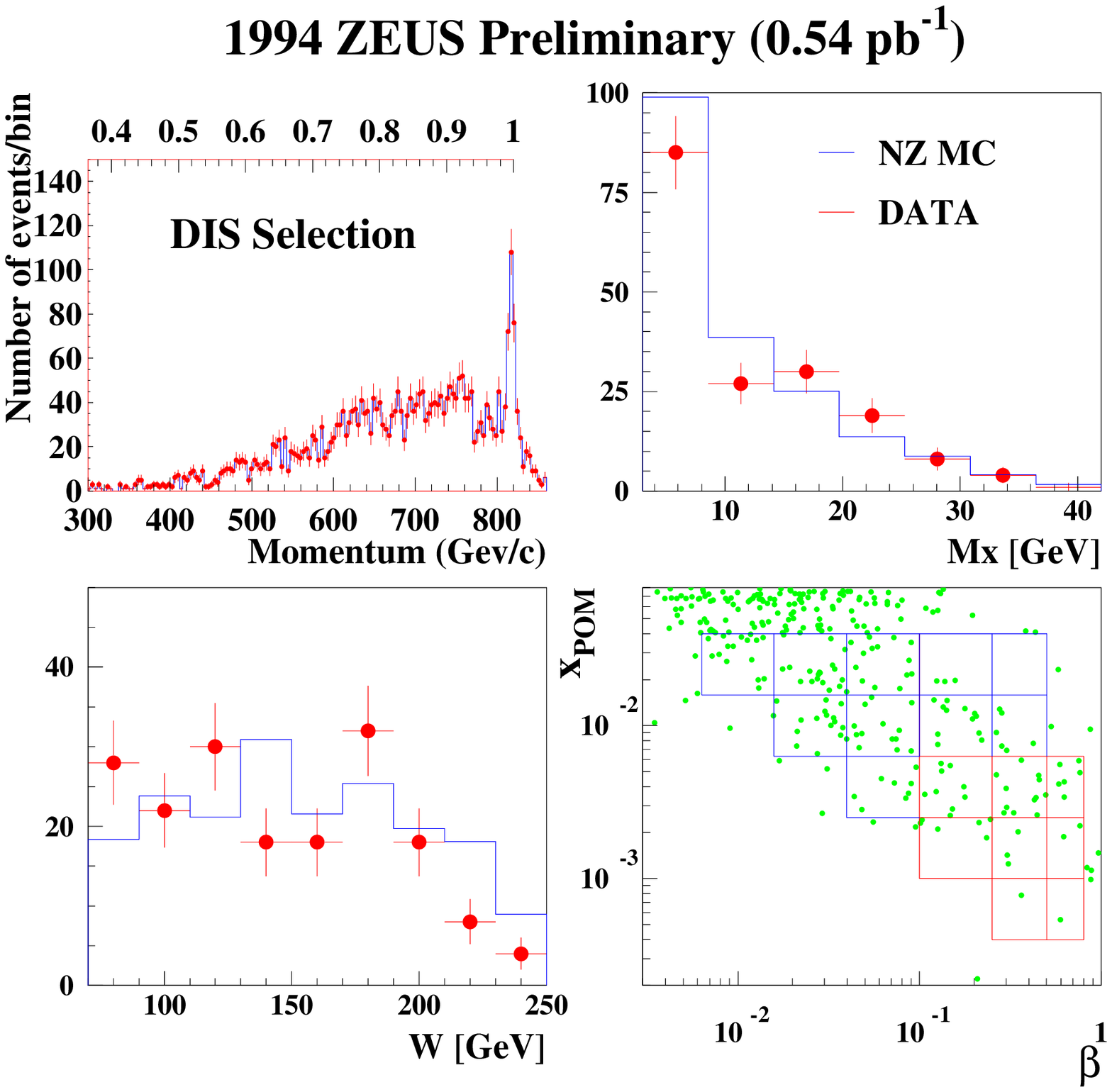}
\end{minipage}
\begin{minipage}[htb]{7cm}
\epsfxsize=6cm
\centering
\leavevmode
\epsfbox{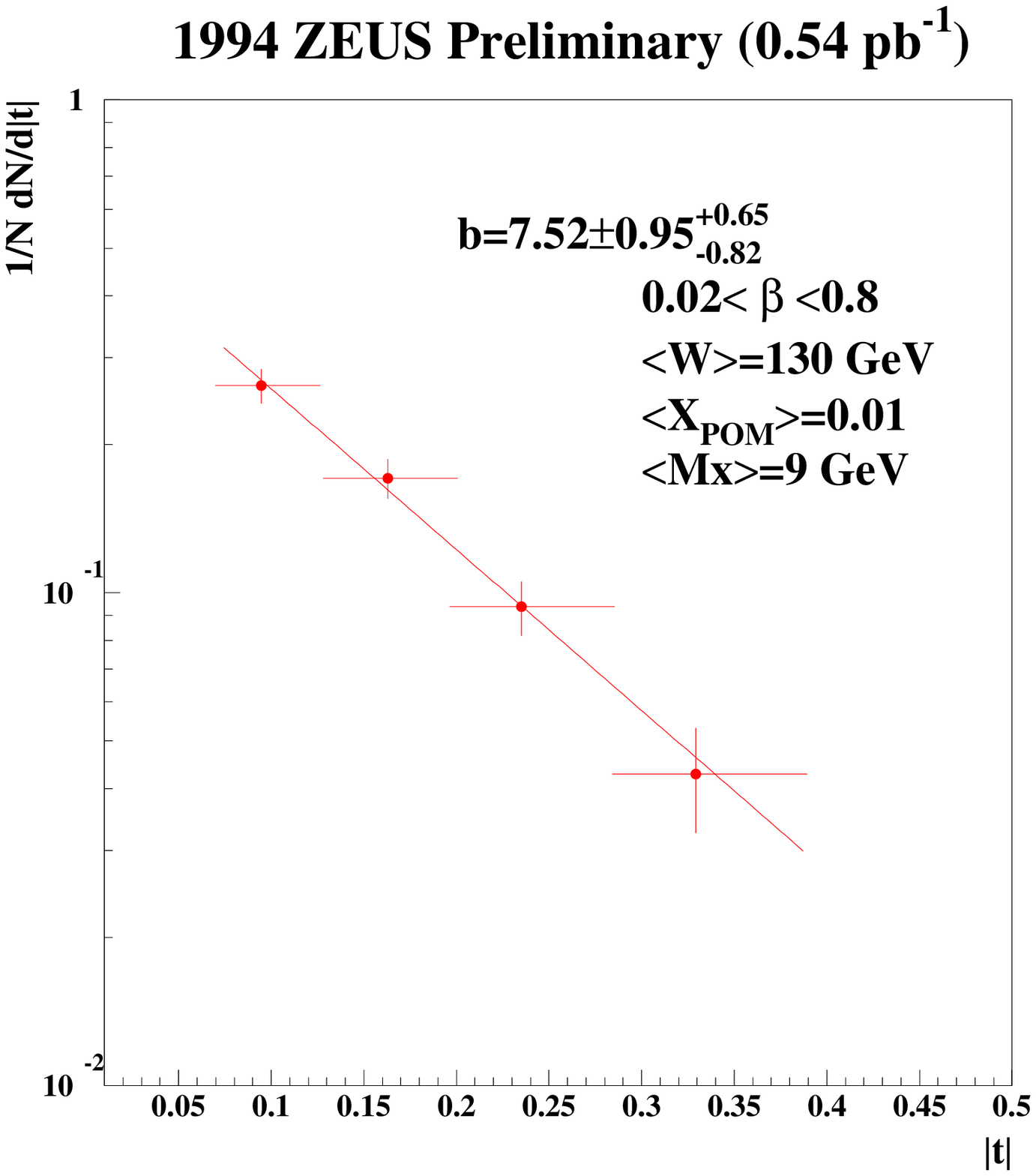}
\end{minipage}
\\
\caption{(a) Observed momentum, $M$, $W$ and $\beta$ versus \xpom\  
distributions in the LPS. The observed $M$ and $W$ distributions are
compared to the NZ Monte Carlo predictions. (b) Corrected $t$ distribution
in the quoted range.}
\end{figure}

\section{Conclusions}

The soft pomeron no longer describes $all$ diffractive data measured at HERA.
As the photon virtuality and/or the vector meson mass increases a new
dependence on $W^2$ emerges.
As we investigate the pomeron more closely, a new type of dynamical pomeron
may begin to play a r\^ole: a dynamical pomeron whose structure is being 
measured in DIS. 
These data are consistent with a partonic description of the exchanged object
which may be described by pQCD.
The experimental work focuses on extending
the lever arms and increasing the precision in $t$, $M^2$, $W^2$ and $Q^2$
in order to explore this new structure. Before more precise tests 
can be made, further theoretical and experimental
input is required to reduce the uncertainties due to 
non-diffractive backgrounds and proton dissociation
as well as the treatment of $F_L$ and radiative corrections.

\section*{Acknowledgements}                                          
                                                                                
The results presented in this talk are a summary of significant 
developments in the study of diffraction at HERA during the last year. 
The financial support of the DESY Directorate and PPARC allowed me to 
participate in this research, whilst based at DESY, for which I am very
grateful.
It is a pleasure to thank Halina Abramowicz, Ela Barberis, Nick Brook, 
Allen Caldwell, John Dainton, Robin Devenish, Thomas Doeker, Robert Klanner, 
Henri Kowalski, Aharon Levy, Julian Phillips, Jeff Rahn, Laurel Sinclair, 
Ian Skillicorn, Ken Smith, Juan Terron, Jim Whitmore and G\"unter Wolf 
for their encouragement, enthusiasm, help and advice.
Finally, thanks to Mike Whalley for his organisation at the workshop
and for keeping me to time in these written proceedings.


\begin{thebibliography}{99}
\bibitem{ingelman} G. Ingelman, J. Phys. G19 (1993) 1633.
\bibitem{goul} K. Goulianos, Phys. Rep. 101 (1983) 169; 
Nucl. Phys. B (Proc. Suppl.) 12 (1990) 110.
\bibitem{dl} 
A. Donnachie and P.V. Landshoff, Nucl. Phys. B244 (1984) 322; 
A. Donnachie, these proceedings.
\bibitem{CDF1} CDF Collab., F. Abe et al., Phys. Rev. D50 (1994) 5550.
\bibitem{Hrho} H1 Collab., S. Aid et al., EPS-0473.
\bibitem{Zrho} ZEUS Collab., M. Derrick et al., DESY 95-143.
\bibitem{Hrho*} H1 Collab., S. Aid et al., EPS-0490.
\bibitem{Zrho*} ZEUS Collab., M. Derrick et al., Phys. Lett. B356 (1995) 601.
\bibitem{Zphi} ZEUS Collab., M. Derrick et al., EPS-0389.
\bibitem{Zphi*} ZEUS Collab., M. Derrick et al., EPS-0397.
\bibitem{Hpsi} H1 Collab., S. Aid et al., EPS-0468.
\bibitem{Zpsi} ZEUS Collab., M. Derrick et al., EPS-0386.
\bibitem{Hpsi*} H1 Collab., S. Aid et al., EPS-0469.
\bibitem{dd} H. Holtmann et al., HEP-PH-9503441. 
\bibitem{CDF2} CDF Collab., F. Abe et al., Phys. Rev. D50 (1994) 5535. 
\bibitem{aston} D. Aston et al., Nucl. Phys. B209 (1982) 56.
\bibitem{schuler} G.A. Schuler and T. Sj\"ostrand, Nucl. Phys. B407 (1993) 539.
\bibitem{levy} A. Levy, DESY 95-204.  
\bibitem{ryskin} M. Ryskin et al., these proceedings.
\bibitem{GRV} M. Gl\"uck, E. Reya and A. Vogt, Phys. Lett. B306 (1993) 391. 
\bibitem{wulff} N. Wulff, University of Hamburg thesis (1994), unpublished.
\bibitem{berger} C. Berger, Proceedings of the Beijing Conference, 
August 1995.
\bibitem{Hd1} H1 Collab., T. Ahmed et al., Phys. Lett. B348 (1995) 681; J.
Dainton, DESY 95-228.
\bibitem{Zd1} ZEUS Collab., M. Derrick et al., Z. Phys. C68 (1995) 569.
\bibitem{ingprytz} G. Ingelman and K. Jansen-Prytz, Z. Phys. C58 (1993) 285.
\bibitem{ariadne} L.~L\"{o}nnblad, Comp. Phys. Comm. 71 (1992) 15.
\bibitem{Zd2} ZEUS Collab., M. Derrick et al., Contribution to the
Beijing Conference, August 1995.
\bibitem{pompyt} P. Bruni and G. Ingelman, DESY 93-187; Proceedings of the 
Europhysics Conference on HEP, Marseille 1993, 595.
\bibitem{nz} N.N. Nikolaev and B.G. Zakharov, Z. Phys. C53 (1992) 331;
M. Genovese, N.N. Nikolaev and B.G. Zakharov, 
KFA-IKP(Th)-1994-37 and {\mbox CERN--TH.13/95}.
\bibitem{buch} W. Buchm\"uller and A. Hebecker, DESY 95-077.
\bibitem{Hd2} H1 Collab., S. Aid et al., EPS-0491; J.P. Phillips, 
Proceedings of the Paris Conference, April 1995.
\bibitem{cf} F. Close and J. Forshaw, HEP-PH-9509251. 
\bibitem{Zjet} ZEUS Collab., M. Derrick et al., Phys. Lett. B356 (1995) 129.
\bibitem{pythia} H.-U. Bengtsson and T. Sj\"ostrand, Comp. Phys. Comm. 
46 (1987) 43; T. Sj\"ostrand, CERN-TH.6488/92.
\bibitem{Zt} ZEUS Collab., M. Derrick et al., DESY 95-194.
\bibitem{D0} D0 Collab., S. Abachi et al., Phys. Rev. Lett. 72 
(1994) 2332; D0 Collab., S. Abachi et al., FERMILAB-PUB-95-302-E (1995). 
\bibitem{CDF3} CDF Collab., F. Abe, et al., Phys. Rev. Lett. 74
(1995) 855.
\bibitem{ada} A. Solano, Ph.D. Thesis, University of Torino 1993 (unpublished);
A. Solano, Nucl. Phys. B (Proc. Suppl.) 25 (1992) 274.
\end{thebibliography}
\end{document}